\documentclass[twoside,fleqn]{article}

\usepackage[dvips]{graphicx}
\usepackage{espcrc2}
\def\ptmin{\hat{p}_T^{\rm min}}
\def\pt{\hat{p}_T}
\def\kt{k_t}
\def\ETJET{E_T^{Jet}}

\newcommand{\AmS}{{\protect\the\textfont2
  A\kern-.1667em\lower.5ex\hbox{M}\kern-.125emS}}

\title{A Global Study of Photon-Induced Jet Production}

\author{J.~M.~Butterworth and \underline{R.~J.~Taylor} 
\\
\bigskip
Department of Physics and Astronomy, University College London,\\ 
Gower Street, London WC1E 6BT, United Kingdom}

\begin{document}

\begin{abstract}
We present results of a global tuning of general purpose Monte Carlo
models to published measurements of $\gamma p \rightarrow$ jets at HERA and
$\gamma\gamma \rightarrow$ jets at LEP and TRISTAN. The principle free
parameters in the tuning are the simulation of the underlying event
and the choice of photon structure.
Several combinations of models are ruled out by the data.
Some consequences of the tuned models at a future linear collider are
discussed.
\end{abstract}

\maketitle

\section{INTRODUCTION}

\begin{figure}[htb]
\vspace{9pt}
{\includegraphics[width=7.5cm]{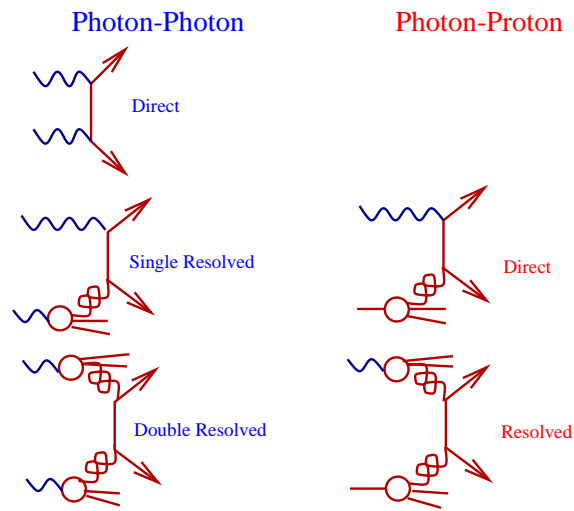}}
\caption{Illustration of the correspondence between the different $\gamma\gamma$ and $\gamma p$ event classes.}
\label{fig:lephera}
\end{figure}

A systematic examination of the manner in which data is described by
Monte Carlo models can give valuable insight into the underlying
physics involved.  However, many such comparisons are to a single
result, limiting the extent to which physics conclusions can be drawn
and raising the possibility that agreement is achieved in one cross
section at the expense of another. The aim of this study is to tune,
for the first time, general purpose Monte Carlo models to the existing
and expanding set of jet photoproduction data. To that end, we
have compared model predictions to published inclusive jet, di-jet and
3-jet data from the ZEUS
\cite{zeus1,zeus2,zeus3,zeus4,zeus5,zeus6,zeus7} and H1 \cite{h11,h12}
experiments at HERA, the OPAL \cite{opal1,opal2} experiment at LEP,
and the TOPAZ \cite{topaz} and AMY \cite{amy} experiments at TRISTAN.
The inclusion of both $\gamma p$ and $\gamma\gamma$ data (see figure
\ref{fig:lephera}) is particularly significant for constraining the
tuning.

Our tuning currently focuses on the role of the so-called
``underlying event'' in hadronic jet production and the extent to
which this can be described by perturbative QCD inspired models. We
constrain the multiparton interaction (minijet) models contained
within the Monte Carlos described below, and use these tunings to
estimate backgrounds at future colliders.

Such a tuning can also lead to a number of practical benefits.  The
method employed allows us to check the consistency of the data
contained in the various publications, which ranges across different
colliders, experiments, years, energies, kinematic regions etc. A
better description of the data by the models can be expected to lead
to a reduction in the systematic errors due to detector corrections in
future measurements.

\section{THE MODELS}

The models tuned were HERWIG 5.9 \cite{hwg}, interfaced to the JIMMY
\cite{mi} eikonal model for multiparton interactions, and PYTHIA 6.125
\cite{pyt}, which contains a new (since version 5.7) mode aiming to
simulate the virtuality of the bremsstrahlung
photons \cite{friberg}. For HERWIG, a previous tuning to DIS data
(CLMAX=5.5, PSPLT=0.65) \cite{distune} was used.  In addition to the
default mode, preliminary investigations were carried out on a
modification to HERWIG, whereby the intrinsic transverse momentum
distribution of the partons in the photon takes the form of a power
law rather than a gaussian. This modified version, referred to as
HERWIG+$\kt$, was motivated by studies at HERA \cite{kt1} and at LEP
\cite{kt2} which indicate that it can lead to a better description of
the data.

\begin{figure}[htb]
\vspace{9pt}
\begin{center}
{\includegraphics[width=4.5cm]{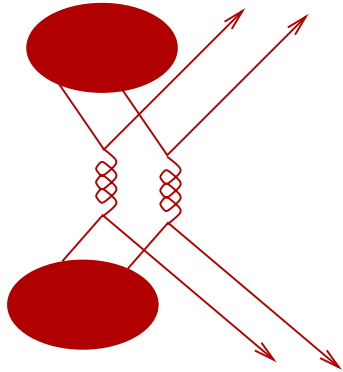}}
\end{center}
\caption{An example of a multiple scattering.}
\label{fig:mi}
\end{figure}

The main free parameters investigated in the tuning are the
description of the ``underlying event'' and the the choice of photon
structure. Multiple hard scatters, illustrated in figure \ref{fig:mi},
can be enabled or not, and it is expected that some of the
distributions studied will exhibit great sensitivity to this
phenomenon. The other facet of the underlying event is the minimum
transverse momentum of the hard scatter(s), hereafter referred to as
$\ptmin$.  Sensitivity to this parameter is greatly increased when
multiple interactions are turned on. As one goes to lower $\pt$, higher
parton densities are being probed, leading to an enhanced probability
that more than one hard scatter will occur. It should also be noted at
this point that, whilst HERWIG 
employs a cluster fragmentation scheme, PYTHIA uses string hadronization.

A number of different parameterisations of the photon structure
function were investigated. Only leading order sets were used since
the matrix elements in the Monte Carlo models are leading order.  The
sets used were GRV 92 \cite{grv}, WHIT2 \cite{whit}, SaS1d and SaS2d
\cite{sas}, and LAC1 \cite{lac}. Throughout this work the GRV 94 LO
\cite{proton} proton pdf was used in the simulation of HERA data.

Finally, the overall normalisation of the Monte Carlo was treated as a
tunable parameter. This is justified within a range of around a factor
of two or less because of the uncertainty in the scale of $\alpha_s$.

\section{FITTING METHOD}

This work was carried out within the framework of a package developed
by the HERA community to permit easy comparison of data to Monte Carlo
\cite{hztool}. Extensions were made to facilitate the inclusion of the
$\gamma\gamma$ data.

The procedure for finding the best fit, for a given set of parameters,
of the Monte Carlo to the data was as follows. An overall $\chi^2$ per
degree of freedom across all the distributions (some 50 in all) was
defined as:
\begin{equation}
  \chi^2 = \frac{1}{n-1}\sum_{i=1}^{n}\frac{(\textrm{MC}(i) - \textrm{Data}(i))^2}{\sigma_{\textrm{\scriptsize{MC}}(i)}^2 + \sigma_{\textrm{\scriptsize{Data}}(i)}^2}
\end{equation}
where Data($i$) and MC($i$) are the values of the distribution in a given bin
$i$ for the data and Monte Carlo respectively. The sum runs over the total 
number of bins $n$ in all the distributions. 

The aim was to minimise this $\chi^2$, and a good fit is when it is
approximately one or less. The normalisation of the Monte Carlo was
varied to find the best fit (The sum in equation (1) is divided by
$n-1$ rather than $n$ to take into account the resulting loss of a
degree of freedom). For the HERA distributions, each data plot was
allowed to float within the quoted correlated error (typically of
15-20\%) and the number of plots was then subtracted from $n$ as
well. Otherwise, all systematic and statistical errors were treated as
uncorrelated and added in quadrature.

The distributions from TRISTAN were not included in the fitting
procedure. There appears to be a discrepancy between the TOPAZ and AMY
results. Indeed, this has been observed in a previous study
\cite{drees}. 

\section{RESULTS AND DISCUSSION}

\begin{table}[hbt]
\newlength{\digitwidth} \settowidth{\digitwidth}{\rm 0}
\catcode`?=\active \def?{\kern\digitwidth}
\caption{HERWIG results (multiparton interactions on, {\bf not} $\kt$ version). The $\chi^2$ per degree of freedom for a particular parameter set is given first for the combined fits (see text), and then for the separate HERA and LEP fits.}
\label{tab:hrw}
\begin{tabular*}{7.5cm}{ccccc}
\hline
$p.d.f.$ & $\ptmin$ & $\chi^2/D.o.F.$ & HERA & LEP \\ 
\hline

WHIT 2 & 1.6 & 3.29 & 4.24 & 1.06 \\
       & 1.8 & 1.89 & 2.27 & 0.99 \\
       & 2.0 & 1.08 & 1.11 & 0.99 \\
       & 2.2 & 1.08 & 1.07 & 1.08 \\
       & 2.4 & 1.47 & 1.60 & 1.14 \\
GRV 92 & 1.6 &      &      & 1.03 \\
       & 1.8 & 1.47 & 1.71 & 0.89 \\
       & 2.0 & 1.35 & 1.46 & 1.09 \\
       & 2.2 &      &      & 1.17 \\
SaS 1d & 1.4 & 2.01 & 2.30 & 1.31 \\
       & 1.6 &      &      & 1.65 \\
       & 1.8 & 1.37 & 1.16 & 1.81 \\
SaS 2d & 1.6 & 1.57 & 1.81 & 0.98 \\
       & 1.8 & 1.28 & 1.40 & 0.97 \\
       & 2.0 &      &      & 1.31 \\
       & 2.2 & 1.27 & 1.25 & 1.31 \\
LAC 1  & 2.2 & 8.70 & 10.54& 4.35 \\
       & 2.6 &      & 2.14 &      \\

\hline
\end{tabular*}
\end{table}

\begin{table}[hbt]
\caption{PYTHIA results (multiparton interactions on). The $\chi^2$ per degree of freedom for a particular parameter set is given first for the combined fits (see text), and then for the separate HERA and LEP fits.}
\label{tab:pyt}
\begin{tabular*}{7.5cm}{ccccc}
\hline
$p.d.f.$ & $\ptmin$ & $\chi^2/D.o.F.$ & HERA & LEP \\ 
\hline
WHIT 2 & 2.4 & 1.17 & 0.91 & 1.86 \\
       & 2.6 & 1.13 & 0.99 & 1.50 \\
       & 2.8 & 1.18 & 1.07 & 1.44 \\
       & 3.0 & 1.24 & 1.12 & 1.52 \\
GRV 92 & 2.4 & 1.21 & 0.96 & 1.89 \\
       & 2.6 & 1.24 & 1.08 & 1.66 \\
       & 2.8 & 1.24 & 1.02 & 1.72 \\
SaS 1d & 1.4 & 1.15 & 1.15 & 1.15 \\
       & 1.6 & 1.15 & 1.19 & 1.03 \\
       & 1.8 & 1.35 & 1.38 & 1.23 \\
SaS 2d & 1.4 & 1.48 & 1.32 & 1.90 \\
       & 1.6 & 1.23 & 1.21 & 1.27 \\
       & 1.8 & 1.09 & 1.08 & 1.09 \\
       & 2.0 & 1.08 & 1.07 & 1.08 \\
       & 2.2 & 1.14 & 1.11 & 1.18 \\
LAC 1  & 2.8 & 6.18 & 5.08 & 10.0 \\

\hline
\end{tabular*}
\end{table}

\begin{figure}[htb]
\vspace{9pt}
{\includegraphics[width=7.5cm]{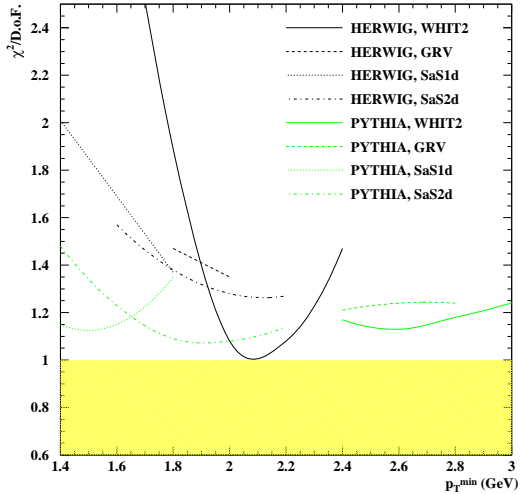}}
\caption{$\chi^2$/DoF for combined LEP and HERA data samples (see text).}
\label{fig:results}
\end{figure}

\begin{figure}[htb]
\vspace{9pt}
{\includegraphics[width=7.5cm]{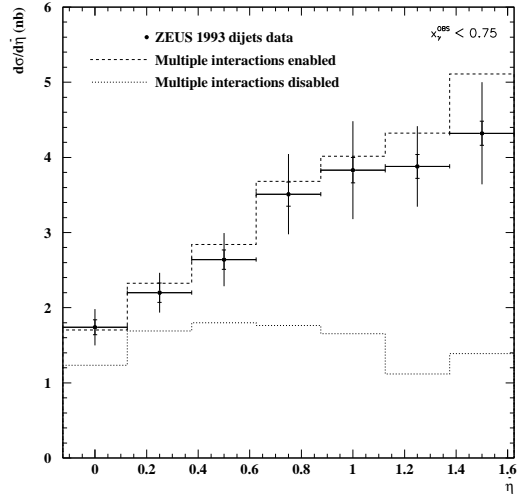}}
\caption{An example of the effect of multiple interactions. The data shown is ZEUS 1993 dijets data [2], along with the HERWIG predictions, using the GRV 92 photon and $\ptmin = 2.0$, with the JIMMY multiple interactions model either enabled or not.}
\label{fig:mi2}
\end{figure}

The results of the fits using HERWIG ({\bf not} the $\kt$ version) are
shown in table \ref{tab:hrw}, and those for PYTHIA in table
\ref{tab:pyt}. Full details of all the fits, including the individual
plots, can be found at
http://www.hep.ucl.ac.uk/\~ ~jmb/HZTOOL/. Figure \ref{fig:results}
shows a summary of the results in graphical form. In most cases there
is a clear, favoured value of $\ptmin$, and overall - at least where
reasonable fits are found - there appears to be a favoured range of
$\sim1.6-2.2$ GeV. Note that although we do not constrain the $\ptmin$
to have the same value for both LEP and HERA fits, when we combine to
give an overall $\chi^2$ we have taken those runs with the same
$\ptmin$. This is motivated by the idea that, as one moves toward low
transverse momentum, two effects can render perturbative QCD
inapplicable.  Not only does $\alpha_s$ become large, but $x$ becomes
small, potentially leading to large $\ln(x)$ corrections. Our
conjecture is that if $\alpha_s$ effects are independent of the centre
of mass energy, then $\ptmin$ can be considered to be a universal
parameter. The low $x$ effects are modelled by multiple interactions,
and so the minijet model is ``universal'', but different 
effects are seen
depending on the cms energy and beam particle type. All the results shown 
have multiple
interactions enabled.  The agreement without multiple interactions is
very poor. A particularly striking example of this is shown in figure
\ref{fig:mi2}, and it is low $\ETJET$ measurements such as this which
are especially valuable for constraining multiple interaction models.

The scaling of the overall normalisation generally agrees between LEP and 
HERA for the good
fits, and is in the range $1.2-1.3$. In addition, there is general
consistency between different datasets for both LEP and HERA. With
regard to pdf sets, WHIT and GRV lead to better fits for HERWIG,
whereas for PYTHIA there are no good fits to LEP data for sets other
than the SaS sets. In no circumstances does it appear possible to
obtain anything close to a good description of the data with LAC1.

The DIS tuned parameters for HERWIG do help, if only marginally, but
there is no firm evidence yet that the HERWIG+$\kt$ modification
improves matters.

\section{LINEAR COLLIDER}

\begin{figure}[htb]
\vspace{9pt}
{\includegraphics[width=7.5cm]{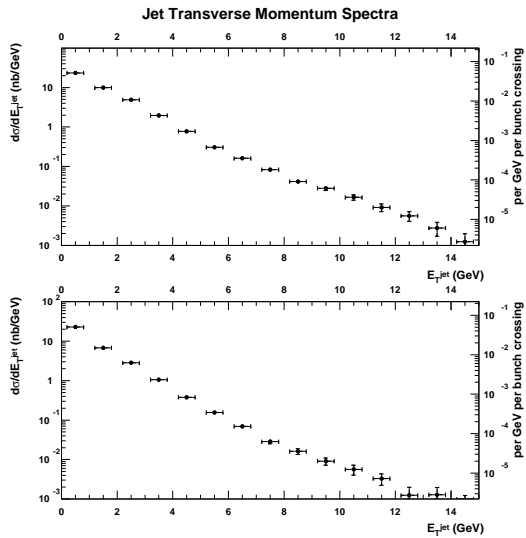}}
\caption{Jet transverse energy differential cross section in the
barrel (top) and endcap (lower) calorimeters of the TESLA conceptual
design report detector \cite{CDR}, for the high luminosity TESLA option.}
\label{fig:lc1}
\end{figure}

The results of the tuning can be used to estimate photon-induced
minijet backgrounds at future colliders. The best fits found using
HERWIG have been extrapolated to linear collider energies ($\sqrt s =
500$ GeV), with the TESLA beamstrahlung and bremsstrahlung spectra
included \cite{circe}. The parameters thus used were the WHIT 2 photon
pdf and a $\ptmin$ of 2.0 GeV with multiple interactions enabled. The
overall normalisation was scaled by a factor of 1.4. The estimate for
the minijet transverse momentum cross-section is shown in figure
\ref{fig:lc1}.

The results are consistent at the 50\% level with independent studies
using PYTHIA and the DELPHI Monte Carlo \cite{sitges}. They indicate
that the backgrounds will not be a concern for detector occupancy and
dosage, but that minijets potentially present a very significant
background to physics owing to the high $\pt$ tail.

\section{CONCLUSIONS}

The general purpose Monte Carlo models HERWIG and PYTHIA have been
tuned to jet photoproduction data from LEP and HERA. An optimal range
of hard scales appears to emerge, and some level of multiparton
interactions is found to be necessary in these models in order that an
adequate description of the data be achieved. Favoured sets of parton
distribution functions are established for each generator but
currently no generator independent conclusion on photon structure is
drawn. The HERWIG tuned parameters are used to estimate photon-induced
backgrounds at the linear collider. This work is ongoing and
up-to-date results can be found on our web page
http://www.hep.ucl.ac.uk/\~ ~jmb/HZTOOL/.

\end{document}